\documentclass[showpacs,preprintnumbers,amsmath,amssymb,aps]{revtex4}

\usepackage{graphicx}
\usepackage{dcolumn}
\usepackage{bm}

\begin{document}

\title{Multichoice Minority Game: Dynamics and Global Cooperation}

\author{F. K. Chow and H. F. Chau}
\affiliation{
  Department of Physics, University of Hong Kong, Pokfulam Road, Hong Kong
}
\date{\today}

\begin{abstract}
In the original two-choice minority game (MG), selfish players cooperate
with each other even though direct communication is not allowed. Moreover,
there is a periodic dynamics in the MG whenever the strategy space size is
much smaller than the number of strategies at play. Do these phenomena
persist if every player has $N_c > 2$ choices where all player's strategies
are picked from a reduced strategy space? We answer this question by
studying a multichoice minority game model known as
MG($N_c$,$|{\mathbb S}|$). Numerical simulation shows that these two models
have very similar global cooperative behaviors. Nevertheless, unlike in the
MG, periodic dynamics does not always appear in the
MG($N_c$,$|{\mathbb S}|$) even when the strategy space size is much smaller
than the number of strategies at play. 
\end{abstract}


\maketitle

\section{Introduction}

In the past decade, there is an explosion of research interest in the study
of complex adaptive system (CAS). In fact, examples of CAS are ubiquitous in
our complex and complicated world. Ecosystems, nervous system, banking
industry, organism and economic systems like stock markets are all examples
of CAS. The study of CAS allowed us to investigate how rich, complicated
pattern structure or behavior emerged from collection of relatively simple
components. 

So, what are the essential features of a complex adaptive system?
According to Levin \cite{CAS}, CAS must have three essential elements:
(i)~sustained diversity and individuality of components, (ii)~localized
interactions among those components and (iii)~an autonomous process that
selects from among those components, based on the results of local
interactions, a subset for replication or enhancement.

In the framework of CAS, econophysicists study the collective behavior of
economic system by building up simple model using statistical mechanical and
nonlinear physical methods. In particular, minority game (MG) \cite{MG1,MG2}
is a simple model of economic system composed of adaptive and inductive
reasoning agents, which successfully captures the minority seeking behavior
in a free market economy. This game was proposed by Challet and Zhang under
the inspiration of the El Farol bar problem introduced by the theoretical
economist Arthur \cite{Elf}.

MG is a toy model of $N$ inductive reasoning players who have to choose one
out of two alternatives independently in each turn by their current best
working strategies or mental model. Those who end up in the minority side
(that is, the choice with the least number of players) win. Although its
rules are remarkably simple, MG shows a surprisingly rich self-organized
collective behavior \cite{MGNu1,MGNu2,MGNu3}. To explain the dynamics of MG,
Hart \emph{et al.} introduced the so-called crowd-anticrowd theory \cite
{MGCAC1,MGCAC2}. Their theory stated that fluctuations arisen in the MG is
controlled by the interplay between crowds of like-minded agents and their
perfectly anti-correlated partners.

Numerical simulation as well as the crowd-anticrowd theory
\cite{MGNu1,MGNu2,MGNu3} showed that the global behavior of MG depends on
three parameters: (i)~the number of players $N$, (ii)~the number of
strategies $S$ each player has and (iii)~the number of the most recent
minority sides $M$ that a strategy depends on. Global cooperation, as
indicated by the fact that average number of players winning the game each
time is larger than that where all players make their choice randomly, is
observed whenever $2^{M+1} \approx N S$ \cite{MGNu1,MGNu2,MGNu3}. Is it true
that global behavior of MG is determined once $N$, $S$ and $M$ are fixed?
Specifically, we ask if it is possible to lock the system in a global
cooperative phase for any fixed values of $N$, $S$ and $M$. We can do so by 
studying the multichoice minority game --- a variant of the MG in which 
players have $N_c > 2$ choices. In our previous work, we have answered the
above question by building up a multichoice minority game model called
MG($N_c$,$N_c^2$) where $N_c$ is the number of player's choice \cite{RMG}. 
In MG($N_c$,$N_c^2$), all strategies are picked from a reduced strategy
space consisting of $N_c^2$ mutually anti-correlated and uncorrelated
strategies only. We found that it is possible to keep (almost) optimal
cooperation amongst the players in almost the entire parameter space since
the MG($N_c$,$N_c^2$) model can be readily extended to MG($N_c$,$N_c^k$)
model for integer $k \in [3,M+1]$.

On the other hand, it was found that the dynamics of the MG is dominated
by a period-$2 \cdot 2^M$ dynamics whenever $2^{M+1} \ll N S$, i.~e.~in the
overcrowding phase \cite{MGNu2,MGP2D,MGTimeS}. In this phase, the strategy
space size is much smaller than the number of strategies at play and thus
each strategy is used by a large number of players. It is such an
overcrowding effect which leads to the period-$2 \cdot 2^M$ dynamics in this
phase. However, will the periodic dynamics also appear in the overcrowding
phase of a multichoice minority game?

In this paper, we would like to continue our previous study by investigating
the global cooperation of players in the multichoice minority game model
known as MG($N_c$,$|{\mathbb S}|$). Besides, we also investigate the
dynamics in the overcrowding phase of such multichoice minority game model.
In Section II, we give a general formalism for constructing the
MG($N_c$,$|{\mathbb S}|$) model. Numerical simulation results on the
dynamics and also global cooperation of our model are reported and
discussed in Section III. Lastly, we sum up by giving a brief conclusion in
Section IV.

\section{The Model}

In this Section, we would like to show how to construct the
MG($N_c$,$|{\mathbb S}|$) model. It is a generalized MG model where all
players have $N_c \ge 2$ choices and they all pick strategies from the
reduced strategy space ${\mathbb S}$ with size $|{\mathbb S}| = N_c^k$ for
integer $k \in [2, M+1]$. (Note that a reduced strategy space is only
consisted of strategies which are significantly different from each other.) 
Here we assume that $N_c$ is a prime power and we label the $N_c$
alternatives as the $N_c$ elements in the finite field $GF(N_c)$. 

In this repeated game, there are $N$ heterogeneous inductive reasoning
players whose aim is to maximize one's own profit from the game. In every
turn, each player has to choose one out of $N_c$ alternatives while direct
communication with others is not allowed. The minority choice, denoted as
$\Omega(t)$ at time $t$, is simply the choice chosen by the least non-zero
number of players in that turn. (In case of a tie, the minority choice is
chosen randomly amongst the choices with the least non-zero number of
players.) The players of the minority choice will gain one unit of wealth
while all the other lose one. The only public information available to all
players is the so-called history $(\Omega(t-M), \ldots, \Omega(t-1))$ which
is the $N_c$-ary string of the minority choice of the last $M$ turns. The 
history can only take on $L \equiv N_c^M$ different states. We label these
states by an index $\mu = 1, \ldots, L$ and denote the history
$(\Omega(t-M), \ldots, \Omega(t-1))$ by the index $\mu(t)$. In fact, players
can only interact indirectly with each other through the history $\mu(t)$.

However, how does each player decide his/her own choice in the game using
inductive reasoning? He/She does so by employing strategies to predict the
next minority choice according to the history where a strategy is a map
sending individual history $\mu$ to the choice $\{0, 1, \ldots, N_c-1\}$. We
represent a strategy $s$ by a vector $\vec{s} \equiv (\chi_s^1, \chi_s^2,
\ldots, \chi_s^L)$ where $\chi_s^\mu$ is the prediction of the minority
choice by the strategy $s$ for the history $\mu$ as illustrated in Table
\ref{tab:t1}.

\begin{table}[ht]
 \caption{The prediction of the minority choice $\chi_s^\mu$ by the strategy
 $\vec{s} \equiv (0,1,2,2,1,0,1,2,0)$ where $N_c = 3$ and $M = 2$.}
 \begin{tabular}{@{\extracolsep{5mm}}cc}
  \hline \hline
  \rule{0pt}{0.15in} history $\mu$ & prediction $\chi_s^\mu$ \\ \hline
  \rule{0pt}{0.15in} (0,0) & 0 \\
  \rule{0pt}{0.15in} (0,1) & 1 \\
  \rule{0pt}{0.15in} (0,2) & 2 \\
  \rule{0pt}{0.15in} (1,0) & 2 \\
  \rule{0pt}{0.15in} (1,1) & 1 \\ 
  \rule{0pt}{0.15in} (1,2) & 0 \\
  \rule{0pt}{0.15in} (2,0) & 1 \\
  \rule{0pt}{0.15in} (2,1) & 2 \\
  \rule{0pt}{0.15in} (2,2) & 0 \\
  \hline \hline
 \end{tabular}
 \label{tab:t1}
\end{table}

In MG($N_c$,$|{\mathbb S}|$), each player is assigned once and for all $S$
randomly drawn strategies from the reduced strategy space ${\mathbb S}$. At
each time step, each player uses his/her own best working strategies to
guess the next minority choice. But how do players decide which strategy is
the best? They use the virtual score, which is the hypothetical profit for a
player using a single strategy throughout the game, to evaluate the
performance of a strategy. The strategy with the highest virtual score is
considered as the best one. 

As mentioned before, all the strategies in MG($N_c$,$|{\mathbb S}|$) are
drawn from the reduced strategy space ${\mathbb S}$. For $|{\mathbb S}| =
N_c^k$ where integer $k \in [2, M+1]$, ${\mathbb S}$ is formed by the
spanning strategies $\vec{v}_a, \vec{v}_u^1, \ldots, \vec{v}_u^{k-1}$ as
follows:
\begin{equation}
 {\mathbb S} = \{ \lambda_a \vec{v}_a + \sum_{i=1}^{k-1}{\lambda_u^i \vec{v}
_u^i} : \lambda_a, \lambda_u^1, \ldots, \lambda_u^{k-1} \in GF(N_c) \}.
\label{E:SS}    
\end{equation}
Note that all arithmetical operations here are performed in the finite
field $GF(N_c)$. Since $N_c$ is a prime power, ${\mathbb S}$ must be a
linear space. Adapted from Ref.~\cite{RMG}, the spanning strategies must
satisfy the following technical conditions:    
\begin{equation}
 v_{ai} \neq 0 \mbox{~for~all~} i, \label{E:Cond_anti} 
\end{equation}
where $v_{ai}$ is the $i$th element of the vector $\vec{v}_a$ and by
regarding $i$ as a uniform random variable between $1$ and $L$,
\begin{equation}
 \begin{array}{c}
  \vspace{2mm} \mbox{Pr} (v_{ui}^1 = j_2 | v_{ai} = j_1) = 1/N_c 
\mbox{~if~Pr} (v_{ai} = j_1) \neq 0 \mbox{~for~all~} j_1, j_2 \in GF(N_c), 
\\
  \mbox{Pr} (v_{ui}^2 = j_3 | v_{ui}^1 = j_2 \mbox{~and~} v_{ai} = j_1) = 
1/N_c \mbox{~if~Pr} (v_{ui}^1 = j_2 \mbox{~and~} v_{ai} = j_1) \neq 0 
\mbox{~for~all~} j_1, j_2, j_3 \in GF(N_c), \\
  \vdots \\
  \begin{split} 
   \mbox{Pr} (v_{ui}^{k-1} = j_k | v_{ui}^{k-2} = & j_{k-1} \mbox{~and~} 
\ldots \mbox{~and~} v_{ui}^1 = j_2 \mbox{~and~} v_{ai} = j_1) = 1/N_c \\ 
   & \mbox{~if~Pr} (v_{ui}^{k-2} = j_{k-1} \mbox{~and~} \ldots \mbox{~and~} 
v_{ui}^1 = j_2 \mbox{~and~} v_{ai} = j_1) \neq 0 \mbox{~for~all~} j_1,
\ldots, j_k \in GF(N_c). 
  \end{split}
 \end{array}
\label{E:Cond_uncorr}
\end{equation}

Under such conditions, we can show that \cite{RMG}
\begin{eqnarray}
& & d( \lambda_{a1} \vec{v}_a + \lambda_{u1}^1 \vec{v}_u^1 + \cdots +
\lambda_{u1}^{k-1} \vec{v}_u^{k-1}, \lambda_{a2} \vec{v}_a + \lambda_{u2}^1
\vec{v}_u^1 + \cdots + \lambda_{u2}^{k-1} \vec{v}_u^{k-1} \nonumber \\
& = & d( [\lambda_{a1} - \lambda_{a2} ] \,\vec{v}_a, [\lambda_{u2}^1 -
 \lambda_{u1}^1 ] \,\vec{v}_u^1 + \cdots + [\lambda_{u2}^{k-1} -
 \lambda_{u1}^{k-1} ] \,\vec{v}_u^k ) \nonumber \\
& = & \left\{ \begin{array}{ll}
 0 & \mbox{if~} \lambda_{a1} = \lambda_{a2} \mbox{~and~} \lambda_{u1}^1 =
\lambda_{u2}^1 \mbox{~and~} \ldots \mbox{~and~} \lambda_{u1}^{k-1} =
\lambda_{u2}^{k-1}, \\
 L & \mbox{if~} \lambda_{a1} \neq \lambda_{a2} \mbox{~and~} \lambda_{u1}^1
= \lambda_{u2}^1 \mbox{~and~} \ldots \mbox{~and~} \lambda_{u1}^{k-1} =
\lambda_{u2}^{k-1}, \\
 L (1-1/N_c) & \mbox{otherwise.}
\end{array} \right. \label{E:HamDis}
\end{eqnarray}

\begin{figure}[ht]
\includegraphics*[scale = 0.5]{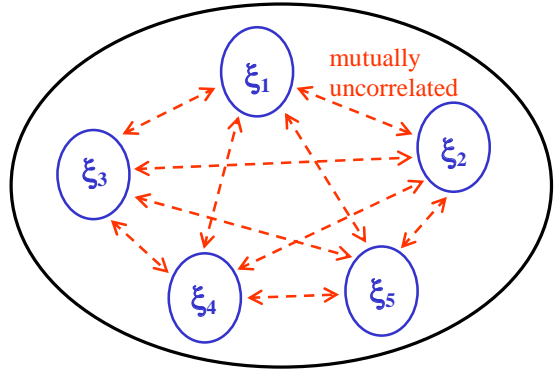}
 \caption{The reduced strategy space ${\mathbb S}$ of MG($N_c$,$N_c^2$) for
$N_c = 5$.}
 \label{fig:f1}
\end{figure}

It is easy to show that the reduced strategy space ${\mathbb S}$ whose size
equals $N_c^k$ is composed of $N_c^{k-1}$ distinct mutually anti-correlated
strategy ensembles (namely, those with same $\lambda_{u}^1, \ldots,
\lambda_{u}^{k-1}$); whereas the strategies of each of these ensembles are
uncorrelated with each other. Indeed, the reduced strategy space
${\mathbb S}$ has a very beautiful structure since any two strategies
randomly drawn from it are either uncorrelated or anti-correlated with each
other \cite{RMG}. Fig.~\ref{fig:f1} shows the structure of the reduced
strategy space ${\mathbb S}$ of MG($N_c$,$N_c^2$) for $N_c = 5$.

\section{Results}

\subsection{Global cooperation of players}

We would like to investigate if there is global cooperation of players in
the MG($N_c$,$|{\mathbb S}|$) model. If the global cooperation does exist,
we want to compare it with that of the minority game. All the results in
this part are taken from 1000 independent runs. In each run, we run 10000
steps starting from initialization before making any measurements. We have
checked that it is already enough for the system to attain equilibrium in
all the cases reported below. Then we took the average values on 15000 steps
after the equilibration.

Let us start by first studying the average attendance
\begin{equation}
 \langle A_i(t) \rangle \equiv \lim_{T \rightarrow 0} \frac{1}{T} \sum_{t} 
A_i(t)
\end{equation}
where the attendance of an alternative $A_i(t)$ is just the number of
players chosen that alternative. Although the attendance of different
alternatives are not the same, their long time averages must be equal since
there is no prior bias for any alternative in this game.
We observed that the average attendance $\langle A_i(t) \rangle$ always
converges to $N/N_c$ in both MG($N_c$,$N_c^2$) and MG($N_c$,$N_c^3$) no
matter what $S$, ${\mathbb S}$ and $M$ are. It is similar to the case of the
MG where the average attendance always converges to $N/2$ \cite{MG1}.

In order to evaluate the performance of players in our model, we study the
mean variance of attendance over all alternatives (or simply the mean
variance)
\begin{equation}
 \Sigma^{2} = \frac{1}{N_c}\sum_{i \in GF(N_c)} [ \langle (A_i(t))^2 \rangle
- \langle A_i(t) \rangle^2] \label{E:var}.
\end{equation}
(We remark that the variance of the attendance of a single alternative was
studied for the MG \cite{MG1}.) In fact, the sum of the variance of all the
attendance represents the total loss of players in the game. The variance
$\Sigma^2$, to first order approximation, is a function of the control
parameter $\alpha$ alone \cite{MGNu2}.

Besides, we will compare the mean variance with that in the random choice
game (RCG) in order to evaluate the significance of the strategies. The RCG
is similar to the MG($N_c$,$|{\mathbb S}|$) model except that all the
players in the RCG make their own decision simply by tossing their own coin.
Indeed, we have showed that the variance is equal to $N(N_c-1)/N_c^2$ for
every alternative in the RCG \cite{MCMG}.

Since the structure of the strategy space ${\mathbb S}$ matches the
assumptions of the crowd-anticrowd theory, we expect the collective behavior
of MG($N_c$,$|{\mathbb S}|$) should follow the predictions of the theory.
Based on our previous work \cite{RMG}, we found that the crowd-anticrowd
theory predicts
\begin{equation}
\Sigma^2 = \left\langle \frac{1}{N_c^{k-1}} \sum_{{\mathbb S}_{\vec{\xi}}}
\sum_{\gamma \in GF(N_c)} \left[ \frac{1}{N_c} \sum_{\eta \neq \gamma}
(N_{\vec{\xi},\gamma} - N_{\vec{\xi},\eta}) \right]^2 \right\rangle,
\label{E:CAC_var}
\end{equation}
where the mutually anti-correlated strategy ensemble
${\mathbb S}_{\vec{\xi}} = \{ \xi_1 \vec{v}_u^1 + \ldots + \xi_{k-1}
\vec{v}_u^{k-1} + \gamma \vec{v}_a : \gamma \in GF(N_c) \}$ for $\vec{\xi}
= (\xi_1, \ldots, \xi_{k-1})$ and $N_{\vec{\xi},\gamma}$ is the number of
players making decision according to the strategy ($\xi_1 \vec{v}_u^1 +
\ldots + \xi_{k-1} \vec{v}_u^{k-1} + \gamma \vec{v}_a$). Note that
$\sum_{{\mathbb S}_{\vec{\xi}}}$ denotes the sum of the variance over all
the mutually anti-correlated strategies ensembles ${\mathbb S}_{\vec{\xi}}$.
We should notice that when averaged over both time and initial choice of
strategies, variance of attendance of different alternatives must be equal
because there is no prior preference for any alternative in our game.

\begin{figure}[ht]
 \includegraphics*[scale = 0.58]{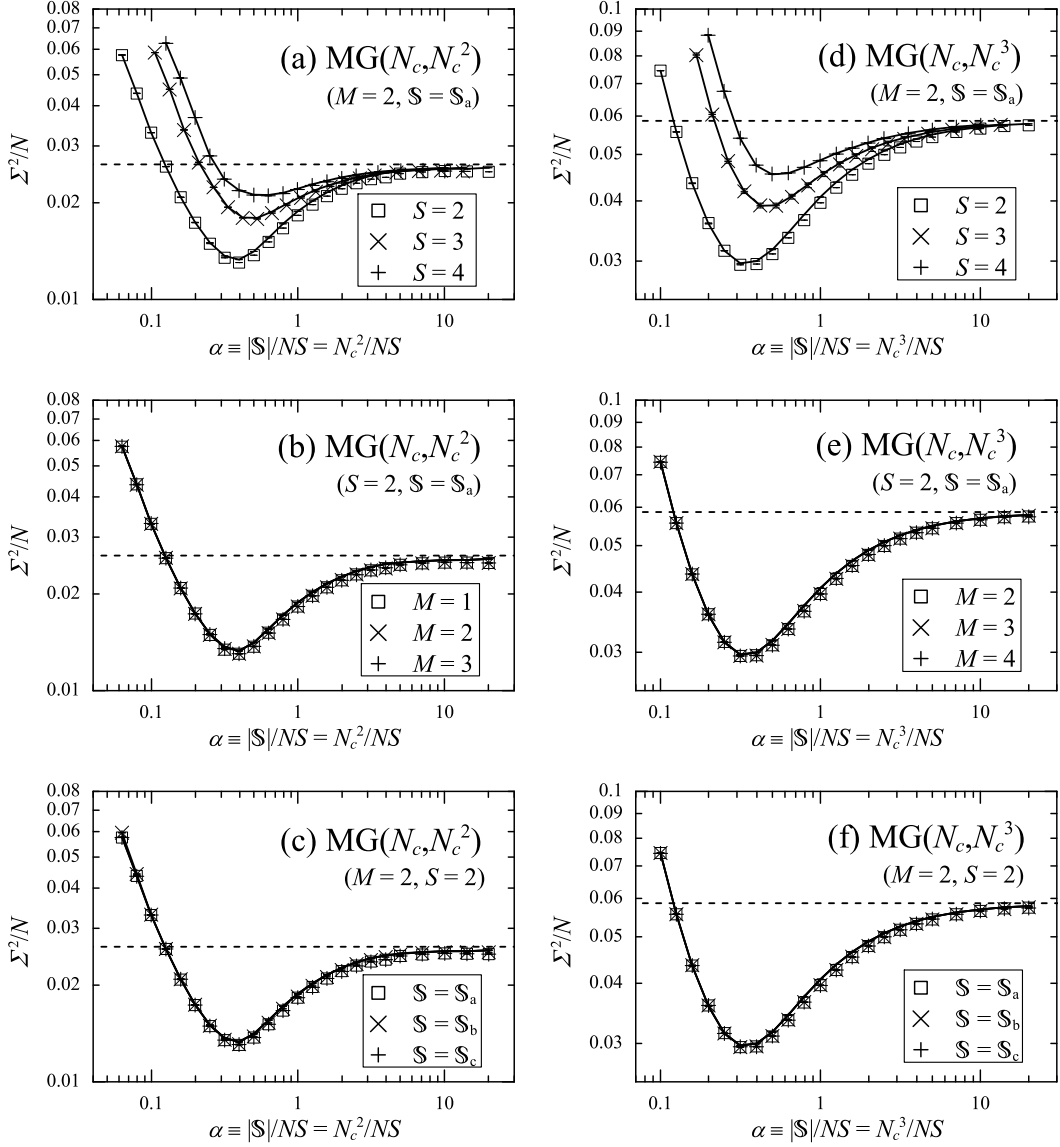}
 \caption{The mean variance $\Sigma^2$ versus the control parameter $\alpha
\equiv |{\mathbb S}|/NS$ in MG($N_c$,$N_c^2$) with $N_c = 37$ and
MG($N_c$,$N_c^3$) with $N_c = 16$. The solid lines are the predictions of
the crowd-anticrowd theory whereas the dashed lines indicate the coin-tossed
value, i.~e.~the corresponding value in the random choice game.}
 \label{fig:f2}
\end{figure}

Fig.~\ref{fig:f2} displays the mean variance of attendance as a function of
the control parameter $\alpha \equiv |{\mathbb S}|/NS$, which is the ratio
of the strategy space size $|{\mathbb S}|$ to the number of strategies at
play $NS$, in the MG($N_c$,$N_c^2$) and MG($N_c$,$N_c^3$) for some typical
$N_c$. For MG($N_c$,$N_c^2$) and MG($N_c$,$N_c^3$), the mean variance of
attendance, $\Sigma^2$, exhibits properties similar to that in the MG
irrespective of $S$, ${\mathbb S}$ and $M$. In particular, the mean variance
$\Sigma^2$ is smaller than the so-called coin-tossed value (i.~e.~the
corresponding value in the RCG) whenever $\alpha \approx 1$. It shows that
players cooperate globally in this parameter range. Moreover, the mean
variance predicted by the crowd-anticrowd theory is consistent with our
numerical finding as shown in Fig.~\ref{fig:f2}.

In both MG($N_c$,$N_c^2$) and MG($N_c$,$N_c^3$), we found that there is an
indication of a second order phase transition. To check if the phase
transition is second order or not, we calculate the order parameter
\cite{RMG,MGOrdP}
\begin{equation}
\theta = \frac{1}{L} \sum_{\mu}{ \left\{ \sum_\Omega{ \left[ \langle
p(\Omega|\mu) \rangle - \frac{1}{N_c} \right]^2 } \right\} }
\end{equation}
where $\langle p(\Omega|\mu) \rangle$ denotes the conditional time
average of the probability for current minority choice $\Omega(t) = \Omega$
given that history $\mu(t) = \mu$. In fact, the order parameter measures the
bias of player's decision to any choice for individual history.

\begin{figure}[ht]
 \includegraphics*[scale = 0.58]{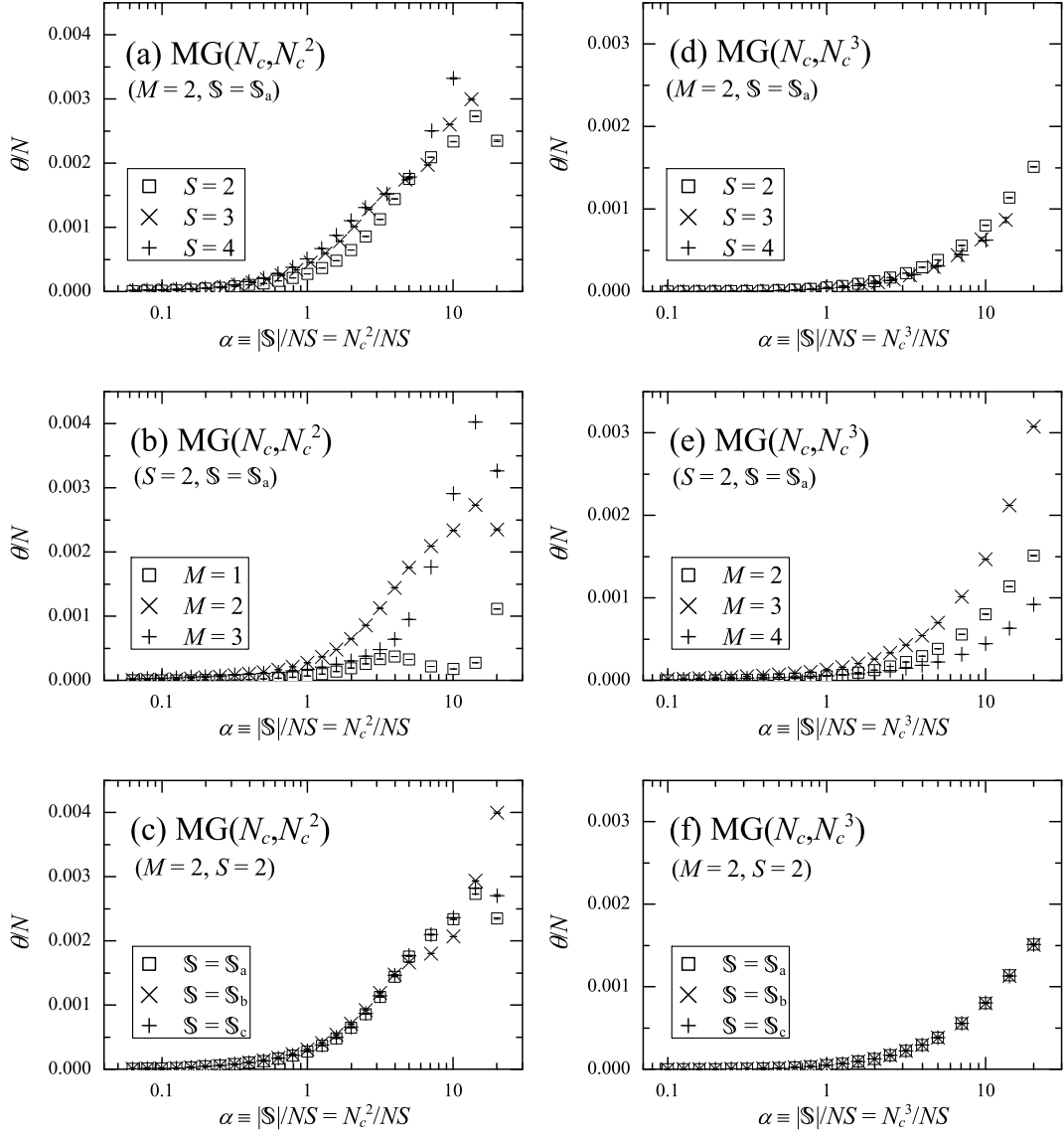}
\caption{The order parameter $\theta$ versus the control parameter $\alpha
\equiv |{\mathbb S}|/NS$ in MG($N_c$,$N_c^2$) with $N_c = 37$ and
MG($N_c$,$N_c^3$) with $N_c = 16$.}
 \label{fig:f3}
\end{figure}

Fig.~\ref{fig:f3} shows that the order parameter always vanishes in
MG($N_c$,$N_c^2$) and MG($N_c$,$N_c^3$) when the control parameter $\alpha$
is smaller than its value corresponding to minimum variance. As a result, we
confirm that the phase transition is a second order one in both
MG($N_c$,$N_c^2$) and MG($N_c$,$N_c^3$).

In summary, our numerical results show that both MG($N_c$,$N_c^2$) and
MG($N_c$,$N_c^3$) model always exhibit global cooperative behavior similar
to that in the original MG. Moreover, all these results agree with the
prediction of the crowd-anticrowd theory. Thus it is reasonable for us to
believe that the MG($N_c$,$|{\mathbb S}|$) model has global cooperative
behavior which agrees with both the original MG and crowd-anticrowd theory.
Therefore, we conclude that we have successfully build up the
MG($N_c$,$|{\mathbb S}|$) model whenever $N_c$ is a prime power.

By using the MG($N_c$,$|{\mathbb S}|$) model, we can always alter the
complexity of each strategy in minority game with fixed $N$, $S$ and $M$
while the cooperative behavior still persists. \emph{As a result, we can
always keep (almost) optimal cooperation amongst the players in almost the
entire parameter space.}

\subsection{Dynamics of the game}

In MG, there is periodic dynamics when $\alpha$ is small, i.~e.~in the
overcrowding phase \cite{MGNu2,MGP2D,MGTimeS}. Thus, it is natural for us to
ask whether such periodic dynamics also exists in the overcrowding phase of
MG($N_c$,$|{\mathbb S}|$).

To investigate the dynamics in MG($N_c$,$|{\mathbb S}|$), we would like to
look at the correlation in the times series of the attendance of an
alternative. Therefore, we study both the conditional and unconditional
autocorrelation functions:
\begin{eqnarray*}
 C_0(i) &=& \frac{\langle A_0(t-i)A_0(t) \rangle - \langle A_0(t-i) \rangle
\langle A_0(t) \rangle}{\langle [A_0(t)]^2 \rangle - \langle A_0(t)
\rangle^2}, \\
 C_0^{\mu}(i) &=& \frac{ \langle {\mathcal A}_0^{\mu}(j-i)
{\mathcal A}_0^{\mu}(j) \rangle - \langle {\mathcal A}_0^{\mu}(j-i) \rangle
\langle {\mathcal A}_0^{\mu}(j)\rangle}{\langle [{\mathcal A}_0^{\mu}(j)]^2
\rangle - \langle {\mathcal A}_0^{\mu}(j) \rangle^2 },
\end{eqnarray*}
where ${\mathcal A}_0^{\mu}(j)$ is the attendance of the room-0 for the
$j$th occurrence of the history $\mu$.
Here we calculate the correlation functions for the room-0 only as the
correlation should be the same for all the alternatives since there is no
prior bias for any alternative in our game.

We observed that both $C_0(i)$ and $C_0^{\mu}(i)$ fluctuate greatly in
different runs of the game with the same configuration. Such phenomenon is
probably due to the incomplete exploration of the history space for
particular initial configurations. Since we are only interested in studying
the generic properties of the autocorrelation functions, we averaged
$C_0(i)$ and $C_0^{\mu}(i)$ over 50 independent runs to eliminate the effect
of the initial configuration. In each run, we took the average values on
$1000L$ steps after running $500L$ steps for equilibrium starting from
initialization.

Before moving further on, let us introduce a concept called the prediction
vector. The prediction vector $\vec{\pi}(\mu)$ of a history $\mu$ is simply
a vector consisted of all the predictions of the minority choice for that
history given by each strategies of the (full/reduced) strategy space used
in MG or MG($N_c$,$|{\mathbb S}|$). Specifically, $\vec{\pi}(\mu) \equiv
(\chi_{\tau_0}^{\mu},\ldots,\chi_{\tau_{|{\mathbb S}|-1}}^{\mu})$ in
MG($N_c$,$|{\mathbb S}|$) where $\tau_j = (\lambda_a \vec{v}_a + \lambda_u^1
\vec{v}_u^1 + \ldots + \lambda_u^{k-1} \vec{v}_u^{k-1})$ for
$j = (\lambda_u^{k-1}N_c^{k-1} + \ldots + \lambda_u^1 N_c + \lambda_a)$.
Table \ref{tab:t2} shows the prediction vectors in the MG($N_c$,$N_c^2$)
with $M = 2$ and $N_c = 3$ for a typical strategy space ${\mathbb S}$.

\begin{table}[ht]
 \caption{The prediction vectors in the MG($N_c$,$N_c^2$) with $M = 2$ and
$N_c = 3$ where the strategy space ${\mathbb S}$ is formed by the spanning
strategies $\vec{v}_a \equiv (1,1,1,1,1,1,1,1,1)$ and $\vec{v}_u^1 \equiv
(0,1,2,0,1,2,0,1,2)$.}
 \begin{tabular}{@{\extracolsep{5mm}}cc}
  \hline \hline
  \rule{0pt}{0.15in} history $\mu$ & prediction vector $\vec{\pi}(\mu)$ \\
\hline
  \rule{0pt}{0.15in} (0,0) & (0,1,2,0,1,2,0,1,2) \\
  \rule{0pt}{0.15in} (0,1) & (0,1,2,1,2,0,2,0,1) \\
  \rule{0pt}{0.15in} (0,2) & (0,1,2,2,0,1,1,2,0) \\
  \rule{0pt}{0.15in} (1,0) & (0,1,2,0,1,2,0,1,2) \\
  \rule{0pt}{0.15in} (1,1) & (0,1,2,1,2,0,2,0,1) \\
  \rule{0pt}{0.15in} (1,2) & (0,1,2,2,0,1,1,2,0) \\
  \rule{0pt}{0.15in} (2,0) & (0,1,2,0,1,2,0,1,2) \\
  \rule{0pt}{0.15in} (2,1) & (0,1,2,1,2,0,2,0,1) \\
  \rule{0pt}{0.15in} (2,2) & (0,1,2,2,0,1,1,2,0) \\
  \hline \hline
 \end{tabular}
 \label{tab:t2}
\end{table}

There is an interesting relationship between the prediction vector and the
history under special circumstances. For some strategy spaces, the
prediction vector of each history $\mu(t) \equiv (\Omega(t-M), \ldots,
\Omega(t-1))$ in MG($N_c$,$N_c^k$) is uniquely determined by one of the
following vectors alone: $\vec{\Lambda}(t-M)$, $\vec{\Lambda}(t-M-1)$,
$\ldots$, $\vec{\Lambda}(t-k+1)$ where $\vec{\Lambda}(t) = (\Omega(t),
\ldots, \Omega(t+k-2))$. For example, suppose the spanning strategies of the
strategy space ${\mathbb S}$ are $\vec{v}_a \equiv (1,1,1,1,1,1,1,1,1)$ and
$\vec{v}_u^1 \equiv (0,0,0,1,1,1,2,2,2)$ in the MG($N_c$,$N_c^2$) with
$M = 2$ and $N_c = 3$. Then the prediction vector are $(0,1,2,0,1,2,0,1,2)$,
$(0,1,2,1,2,0,2,0,1)$ and $(0,1,2,2,0,1,1,2,0)$ for histories in which
$\Omega(t-2)$ equals $0$, $1$ and $2$ respectively. In this case, the
prediction vector of an arbitrary history $\mu(t)$ is uniquely determined
by $\Omega(t-2)$.

We examined the behavior of $C_0^{\mu}(i)$ in the overcrowding phase for
MG($N_c$,$N_c^2$) and MG($N_c$,$N_c^3$) with $2 \le M \le 4$ and various
strategy spaces. We noticed that $C_0^{\mu}(i)$ exhibits periodic features
only in MG($N_c$,$N_c^3$) with $M = 2$ no matter what strategy space
${\mathbb S}$ is used. Such observation indicates periodic dynamics of
individual history persists in the MG($N_c$,$N_c^3$) game with $M = 2$. For
other studied games, the values of $C_0^{\mu}(i>1)$ are very small and
sometimes even $C_0^{\mu}(1)$ is nearly zero. These observations suggest
only the dependence between consecutive ${\mathcal A}_0^{\mu}(j)$ is
non-negligible in these MG($N_c$,$|{\mathbb S}|$) games.

\begin{figure}[ht]
 \includegraphics*[scale = 0.52]{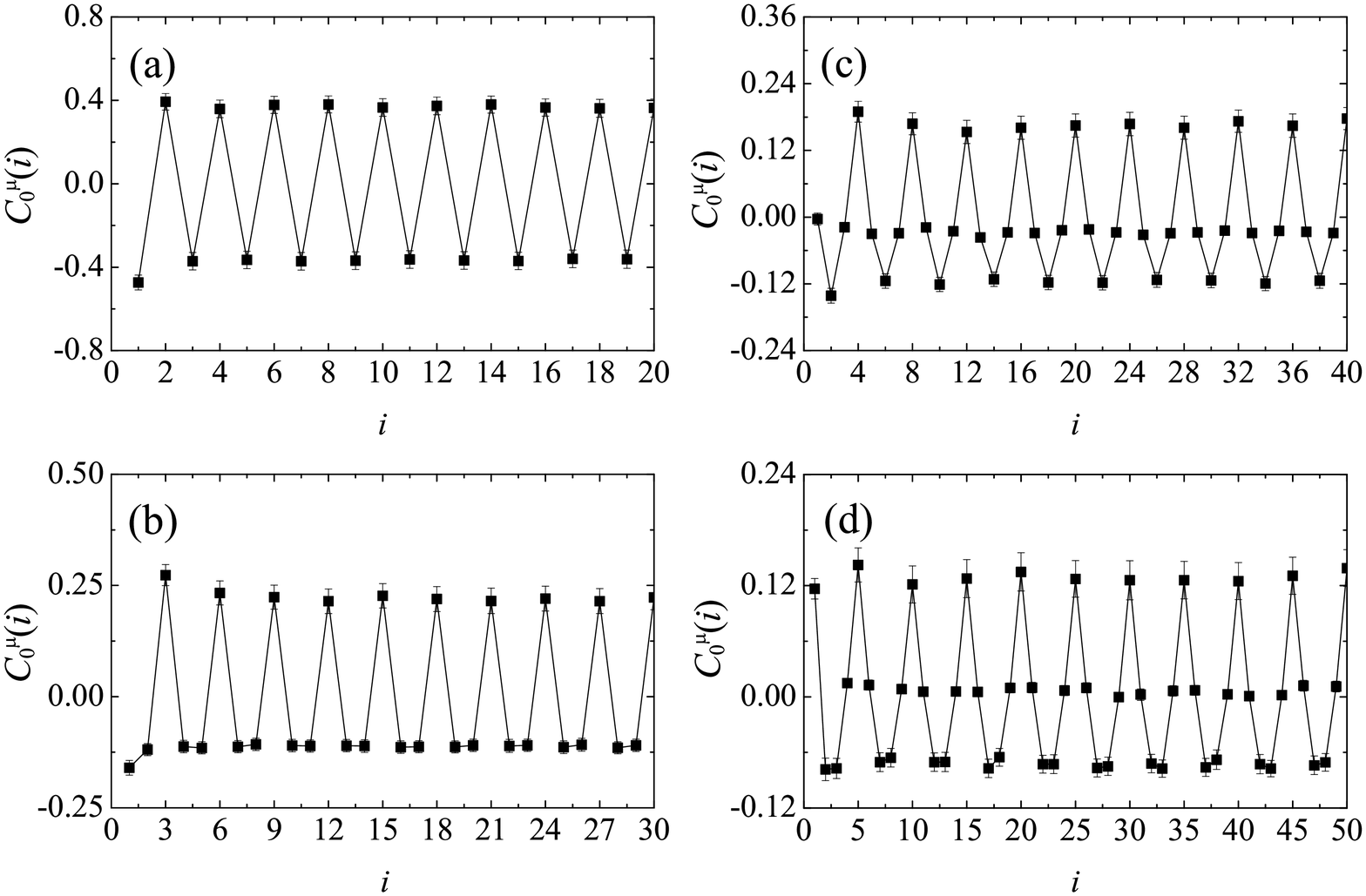}
 \caption{The conditional autocorrelation function $C_0^{\mu}(i)$ versus $i$
in $N_c$-choice MG with $S = 2$, $M = 2$ and $\alpha \approx 0.06$ where
(a)~$N_c = 2$, (b)~$N_c = 3$, (c)~$N_c = 4$, (d)~$N_c = 5$. Our numerical
results show that $C_0^{\mu}(i)$ exhibits similar behavior for all other
histories and $M = 3, 4$.}
 \label{fig:f4}
\end{figure}

On the contrary, we observed that $C_0^{\mu}(i)$ is a period-2 function of
$i$ in the overcrowding phase of the 2-choice MG for any history $\mu$. It
originates from the persistence of a period-2 dynamics for every possible
history in this game \cite{MGP2D}. In the overcrowding phase, there is a
high probability for a player to employ similar strategies since the number
of strategies at play is much larger than the strategy space size. As a
result, for odd occurence of each history, every player has the same
probability for choosing any alternative which results in $A_0(\mbox{odd }t)
\approx N/2$. Moreover, on the even occurence of each history, every players
must choose the alternative which is the minority choice of the last odd
occurence of the same history and thus $A_0(\mbox{even }t)$ is either $\ll
N/2$ or $\gg N/2$. If this periodic dynamics can persist, $C_0^{\mu}(i)$
must be a period-2 function of $i$ in the overcrowding phase of the 2-choice
MG for any history $\mu$. In fact, there is a similar period-$N_c$ dynamics
for each history of the $N_c$-choice MG with $N_c > 2$ in which the full
strategy space is used (see fig.~\ref{fig:f4}).

We expect there is period-$N_c$ dynamics for every possible history in the
overcrowding phase of MG($N_c$,$|{\mathbb S}|$) due to its similarity with
the MG. However, why the period-$N_c$ dynamics cannot always persist in our
game? Such phenomenon is resulted from the ``interference'' of the periodic
dynamics of individual history. In MG($N_c$,$N_c^k$), each prediction vector
corresponds to more than one history if $M > (k-1)$ since the prediction
vector can only take on $N_c^{k-1}$ to $(N_c-1)N_c^{k-1}$ different states
for $|{\mathbb S}| = N_c^k$. Moreover, the virtual scores of all player's
strategies must change in the same way for those histories with the same
prediction vector. As a result, the periodic dynamics of individual history
must ``interfere'' with those corresponding to the same prediction vector
and thus the periodic dynamics cannot persist. In addition, there is a
one-to-one correspondence between the prediction vector and history for both
the (full/reduced) strategy spaces used in MG and MG($N_c$,$N_c^k$) with
$M = k - 1$. Consequently, the periodic dynamics of individual history
always persist in these games since they no longer ``interfere'' with each
other. 
 
\begin{figure}[ht]
 \includegraphics*[scale = 0.52]{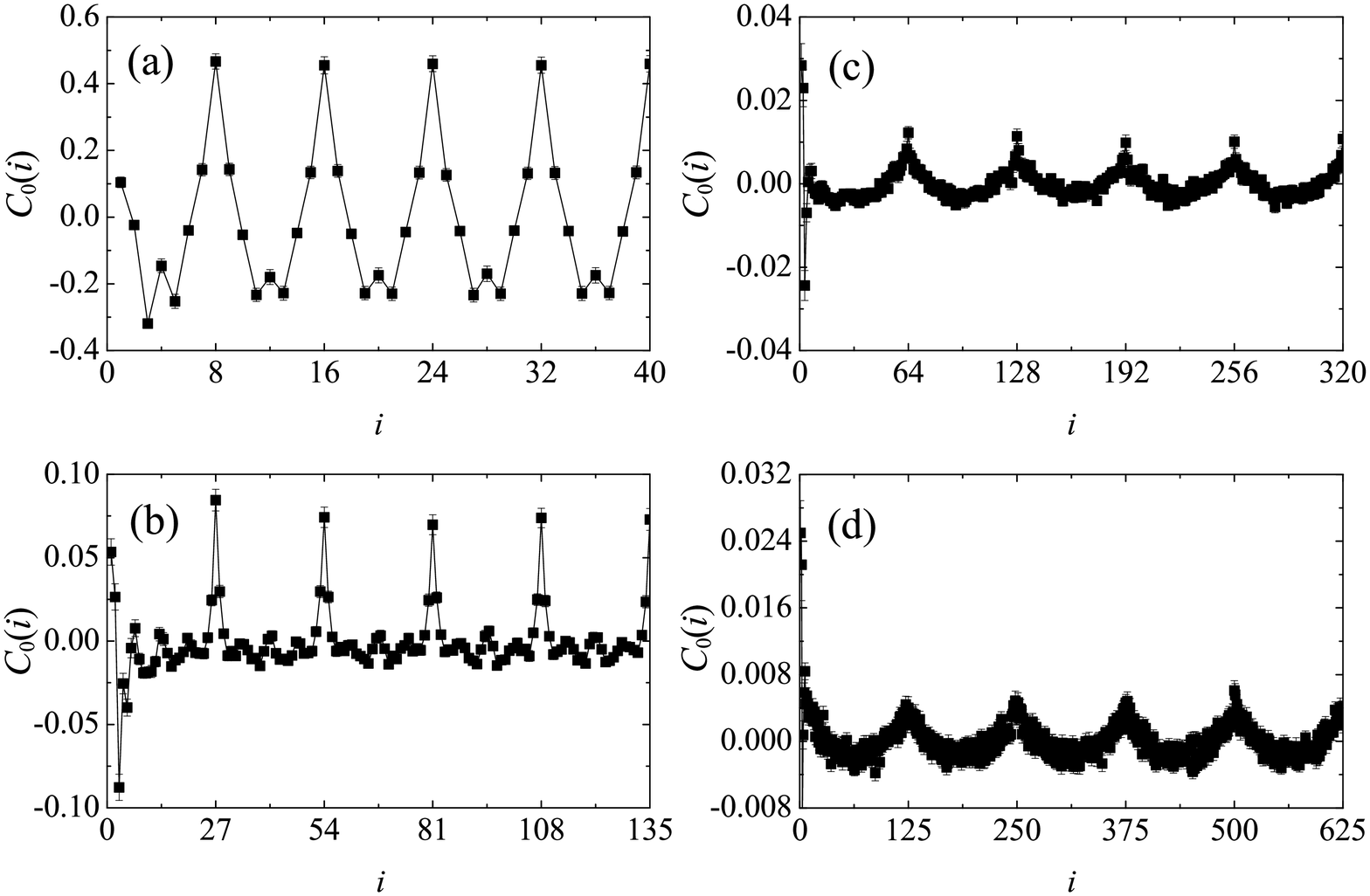}
 \caption{The autocorrelation function $C_0(i)$ versus $i$ in $N_c$-choice
MG with $S = 2$, $M = 2$ and $\alpha \approx 0.06$ where (a)~$N_c = 2$,
(b)~$N_c = 3$, (c)~$N_c = 4$, (d)~$N_c = 5$. Our numerical results show that
$C_0(i)$ exhibits similar behavior for $M = 3$ and $4$.}
 \label{fig:f5}
\end{figure}

On the other hand, is it possible to have any periodic dynamics in the
system of MG and MG($N_c$,$|{\mathbb S}|$)? Zheng and Wang revealed that
period-$2\cdot 2^M$ multi-peak structure appears in the unconditional
autocorrelation function of the logarithm change of the attendance of an
alternative in the 2-choice MG when $\alpha$ is small \cite{MGTimeS}. Thus
all players must have similar behavior for every $2\cdot 2^M$ time steps in
such case. In other words, there is always a period-$2\cdot 2^M$ dynamics in
the overcrowding phase of the 2-choice MG. 
Fig.~\ref{fig:f5} showed that a period-$N_c\cdot N_c^M$ dynamics also exists
in the overcrowding phase of the $N_c$-choice MG for $N_c > 2$ although it
is weaker than the periodic dynamics in the 2-choice MG.

\begin{figure}[ht]
 \includegraphics*[scale = 0.48]{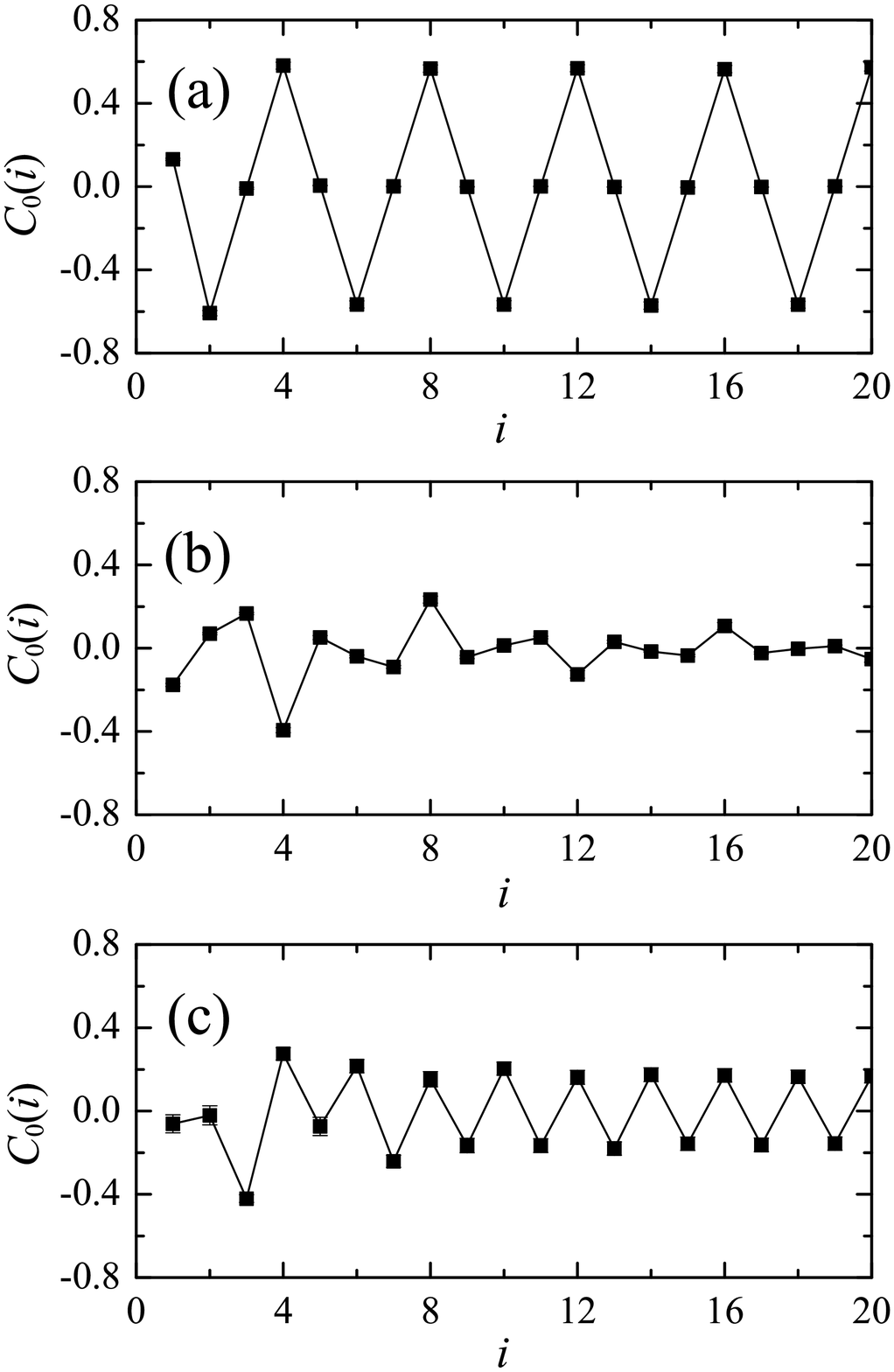}
 \caption{The autocorrelation function $C_0(i)$ versus $i$ in
MG($N_c$,$N_c^2$) with $N_c = 2$, $S = 2$, $M = 2$ and $\alpha \approx
0.06$ where the strategy space ${\mathbb S}$ is formed by the spanning
strategies $\vec{v}_a = (1,1,1,1)$ and (a)~$\vec{v}_u^1 = (0,1,0,1)$,
(b)~$\vec{v}_u^1 = (0,0,1,1)$, (c)~$\vec{v}_u^1 = (0,1,1,0)$.}
 \label{fig:f6}
\end{figure}

\begin{figure}[ht]
 \includegraphics*[scale = 0.52]{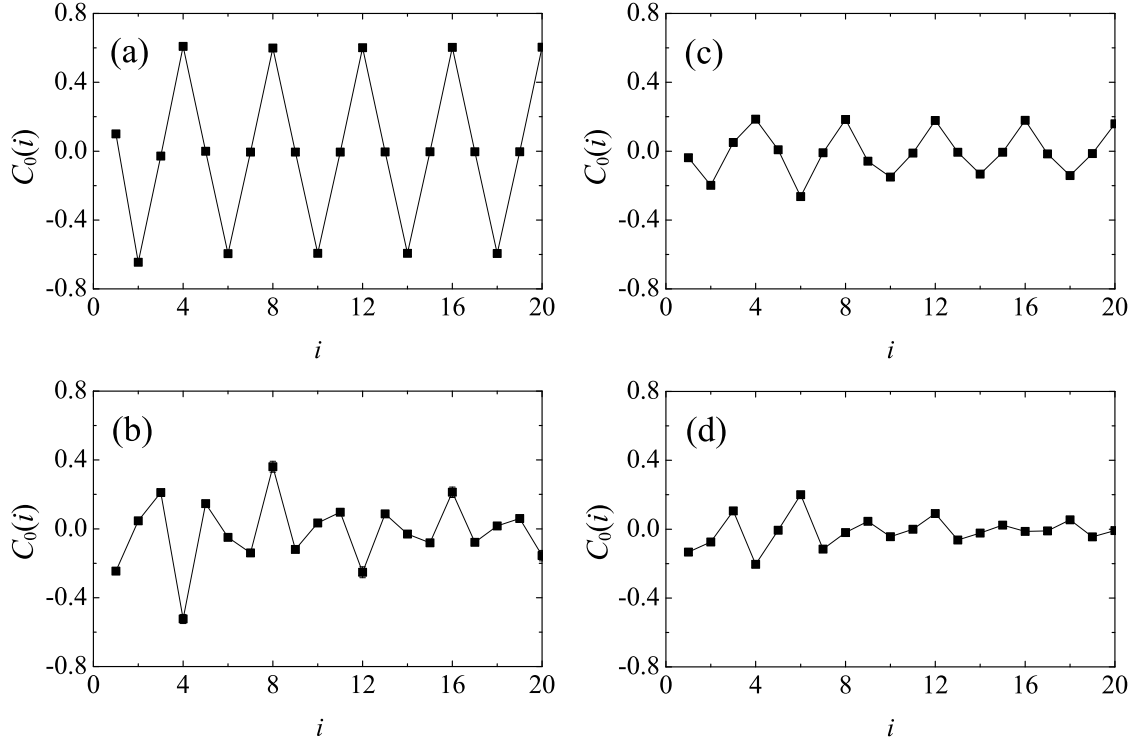}
 \caption{The autocorrelation function $C_0(i)$ versus $i$ in
MG($N_c$,$N_c^2$) with $N_c = 2$, $S = 2$, $M = 3$ and $\alpha \approx
0.06$ where the strategy space ${\mathbb S}$ is formed by the spanning
strategies $\vec{v}_a = (1,\ldots,1)$ and (a)~$\vec{v}_u^1 = (0,1)^4$, (b)~
$\vec{v}_u^1 = (0,0,1,1)^2$, (c)~$\vec{v}_u^1 = (0,\ldots,0,1,\ldots,1)$,
(d)~irregular $\vec{v}_u^1$. Note that the length of a strategy $L = N_c^M$
and $(\sigma)^n$ denotes the vector of $n$ consecutive segment $\sigma$.}
 \label{fig:f7}
\end{figure}

\begin{figure}[ht]
 \includegraphics*[scale = 0.52]{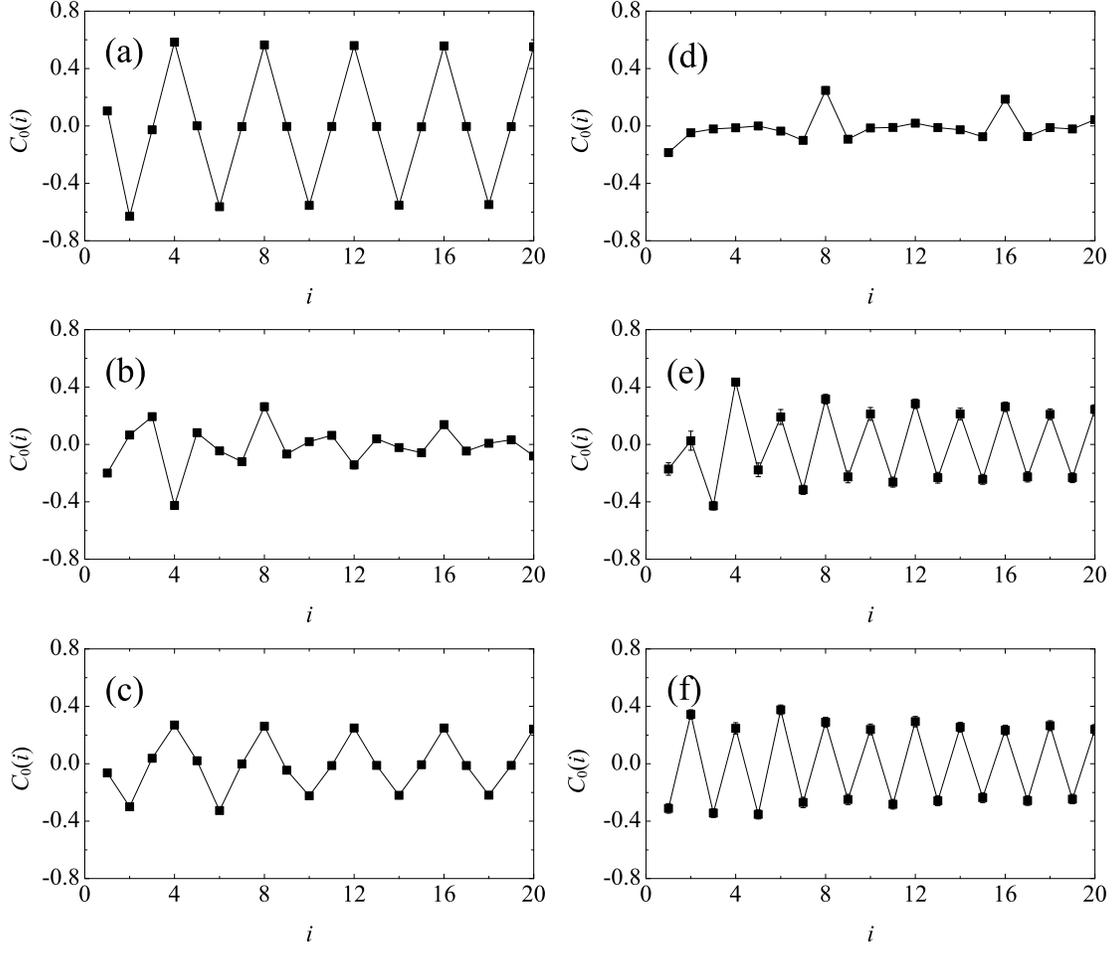}
 \caption{The autocorrelation function $C_0(i)$ versus $i$ in
MG($N_c$,$N_c^2$) with $N_c = 2$, $S = 2$, $M = 4$ and $\alpha \approx 
0.06$ where the strategy space ${\mathbb S}$ is formed by the spanning
strategies $\vec{v}_a = (1,\ldots,1)$ and (a)~$\vec{v}_u^1 = (0,1)^8$, (b)~
$\vec{v}_u^1 = (0,0,1,1)^4$, (c)~$\vec{v}_u^1 = (0,\ldots,0,1,\ldots,1)^2$,
(d)~$\vec{v}_u^1 = (0,\ldots,0,1,\ldots,1)$,
(e)~$\vec{v}_u^1 = (0,1,1,0)^4$, (f)~irregular $\vec{v}_u^1$ with the same
notation as in Fig.\ref{fig:f7}.}
 \label{fig:f8}
\end{figure}

\begin{figure}[ht]
 \includegraphics*[scale = 0.52]{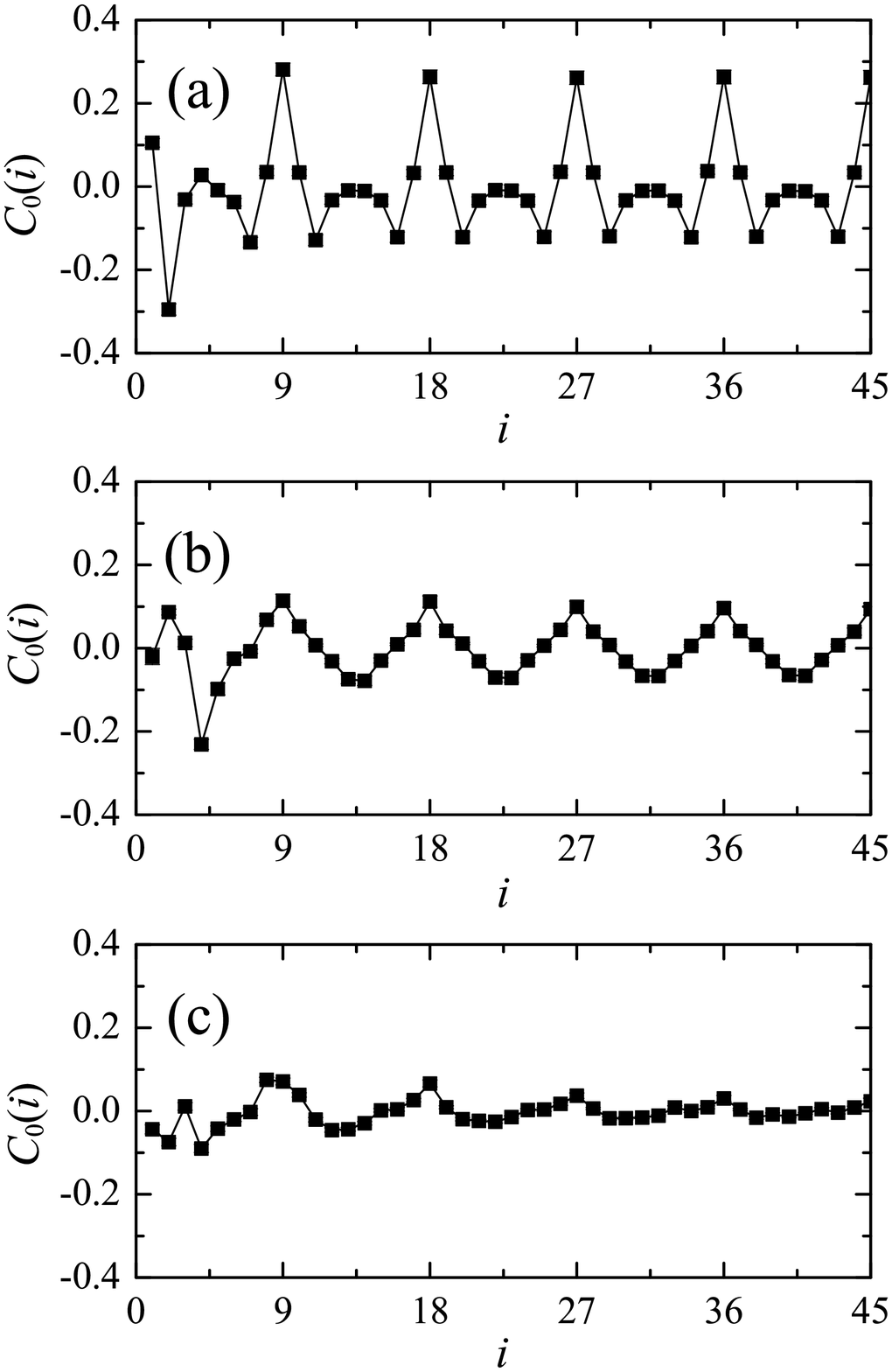}
 \caption{The autocorrelation function $C_{0,0}(i)$ versus $i$ in
MG($N_c$,$N_c^2$) with $N_c = 3$, $S = 2$, $M = 2$ and $\alpha \approx
0.06$ where the strategy space ${\mathbb S}$ is formed by the spanning
strategies $\vec{v}_a = (1,\ldots,1)$ and (a)~$\vec{v}_u^1 = (0,1,2)^3$, 
(b)~$\vec{v}_u^1 = (0,0,0,1,1,1,2,2,2)$, (c)~irregular $\vec{v}_u^1$ with
the same notation as in Fig.\ref{fig:f7}.}
 \label{fig:f9}
\end{figure}

\begin{figure}[ht]
 \includegraphics*[scale = 0.52]{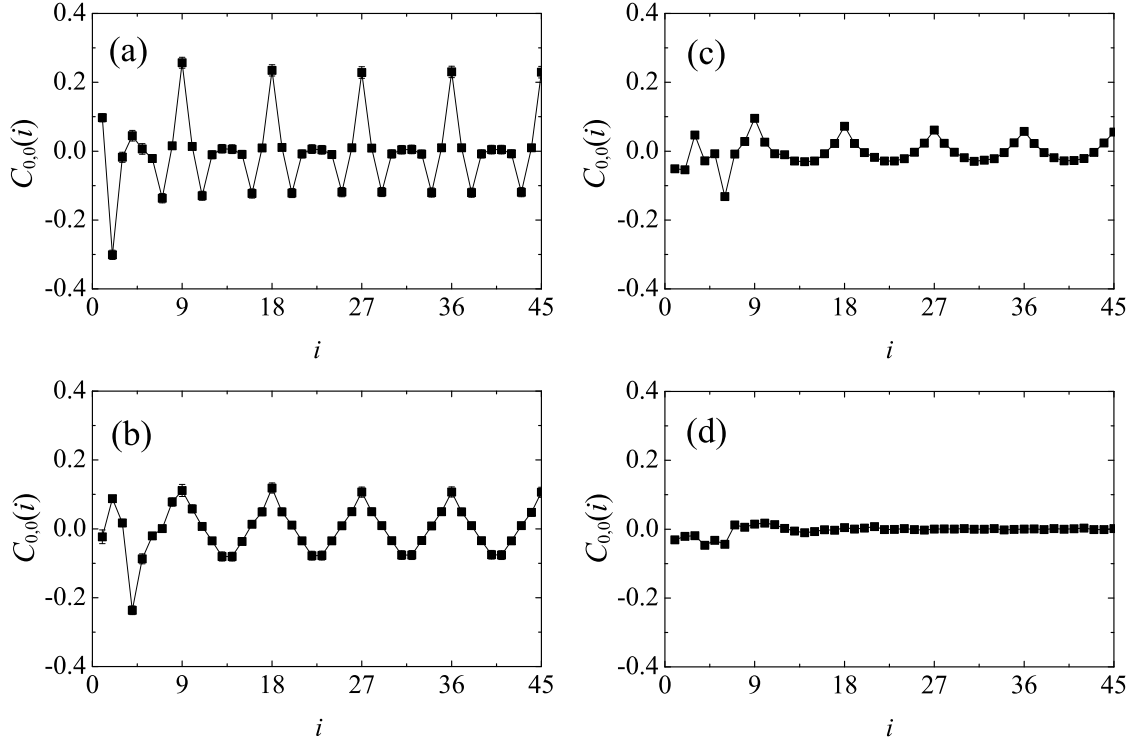}
 \caption{The autocorrelation function $C_0(i)$ versus $i$ in
MG($N_c$,$N_c^2$) with $N_c = 3$, $S = 2$, $M = 3$ and $\alpha \approx
0.06$ where the strategy space ${\mathbb S}$ is formed by the spanning
strategies $\vec{v}_a = (1,\ldots,1)$ and (a)~$\vec{v}_u^1 = (0,1,2)^9$,
(b)~$\vec{v}_u^1 = (0,0,0,1,1,1,2,2,2)^3$,
(c)~$\vec{v}_u^1 = (0,\ldots,0,1,\ldots,1,2\ldots,2)$, (d)~irregular
$\vec{v}_u^1$ with the same notation as in Fig.\ref{fig:f7}.}
 \label{fig:f10}
\end{figure}

\begin{figure}[ht]
 \includegraphics*[scale = 0.52]{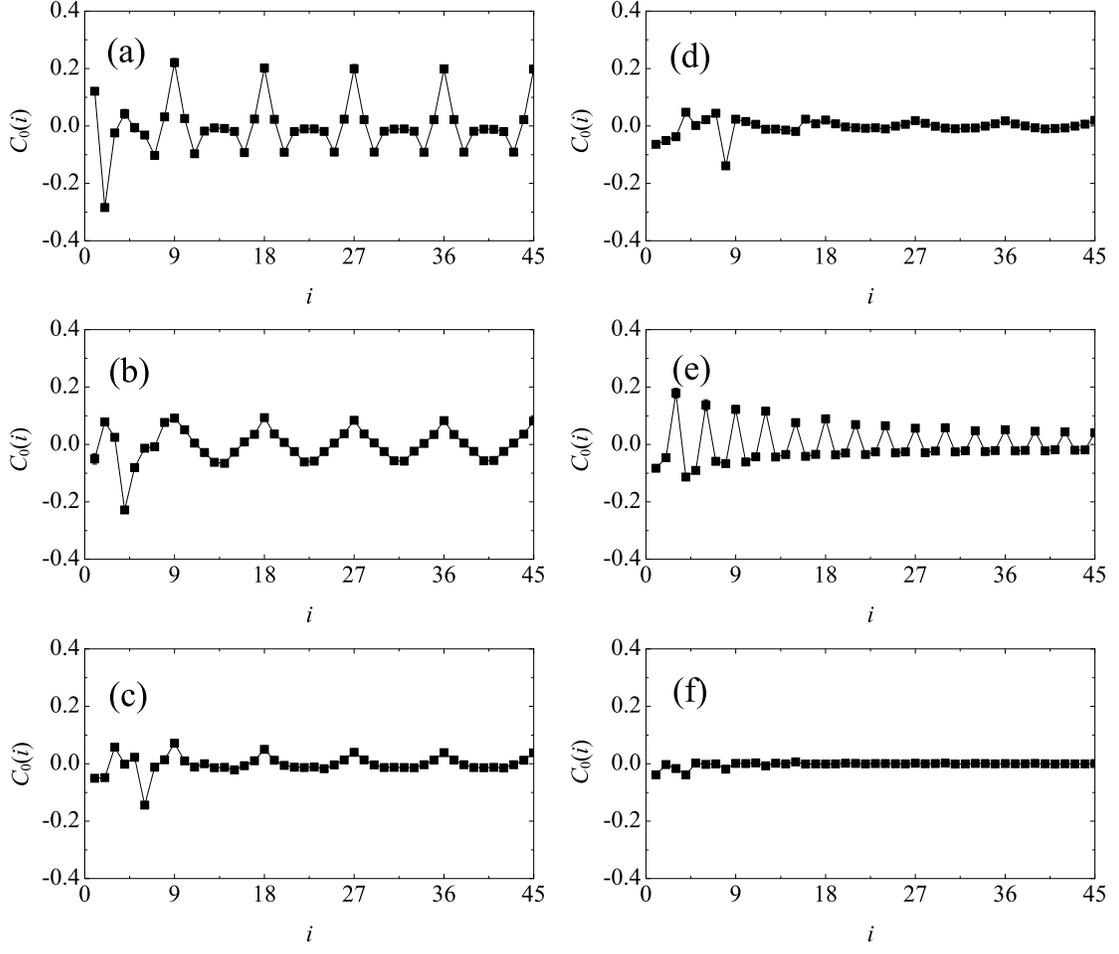}
 \caption{The autocorrelation function $C_0(i)$ versus $i$ in 
MG($N_c$,$N_c^2$) with $N_c = 3$, $S = 2$, $M = 4$ and $\alpha \approx 
0.06$ where the strategy space ${\mathbb S}$ is formed by the spanning
strategies $\vec{v}_a = (1,\ldots,1)$ and (a)~$\vec{v}_u^1 = (0,1,2)^{27}$,
(b)~$\vec{v}_u^1 = (0,0,0,1,1,1,2,2,2)^9$,
(c)~$\vec{v}_u^1 = (0,\ldots,0,1,\ldots,1,2\ldots,2)^3$,
(d)~$\vec{v}_u^1 = (0,\ldots,0,1,\ldots,1,2\ldots,2)$,
(e)~$\vec{v}_u^1 = (0,1,2,2,0,1,1,2,0)^9$, (f)~irregular $\vec{v}_u^1$ with
the same notation as in Fig.\ref{fig:f7}.}
 \label{fig:f11}
\end{figure}

\begin{figure}[ht]
 \includegraphics*[scale = 0.52]{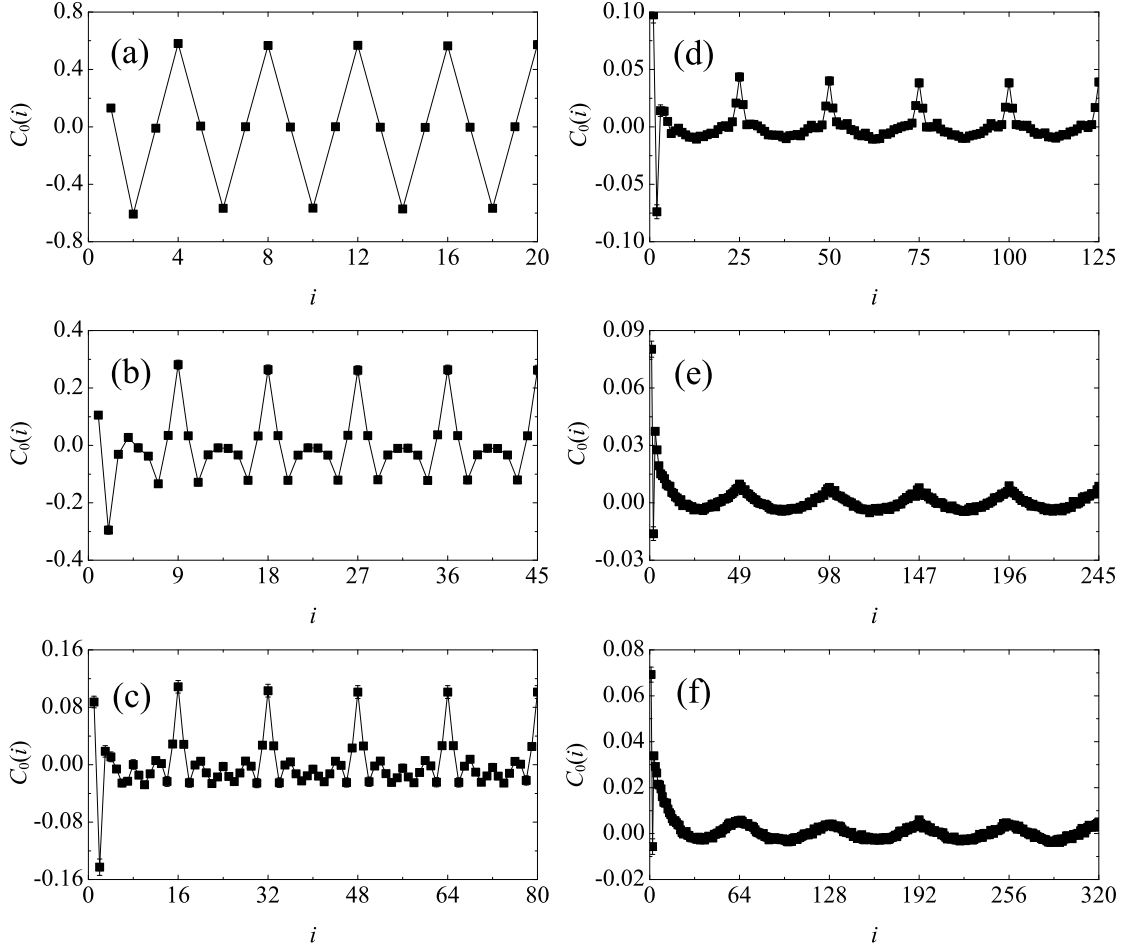}
 \caption{The autocorrelation function $C_0(i)$ versus $i$ in
MG($N_c$,$N_c^2$) with $S = 2$, $M = 2$ and $\alpha \approx 0.06$ for the
strategy space ${\mathbb S}$ formed by the spanning strategies $\vec{v}_a 
= (1,\ldots,1)$ and $\vec{v}_u^1 = (0,1,\ldots,N_c-1)^{L/N_c}$ where 
(a)~$N_c = 2$, (b)~$N_c = 3$, (c)~$N_c = 4$, (d)~$N_c = 5$, (e)~$N_c = 7$,
(f)~$N_c = 8$ with the same notation as in Fig.\ref{fig:f7}. Our numerical
results show that $C_0(i)$ exhibits similar periodic behavior in
MG($N_c$,$N_c^2$) with the same settings but $M = 3$ and $4$.}
 \label{fig:f12}
\end{figure}

In order to study the dynamics in MG($N_c$,$|{\mathbb S}|$), we investigated
the properties of $C_0(i)$ in MG($N_c$,$N_c^2$) and MG($N_c$,$N_c^3$).
Figs.~\ref{fig:f6} to \ref{fig:f12} display $C_0(i)$ as a function of $i$
for MG($N_c$,$N_c^2$) with different parameters. We found that the dynamics
of the system in the overcrowding phase of MG($N_c$,$|{\mathbb S}|$) is
remarkably different from that of MG.
First, $C_0(i)$ does not always exhibit periodic features in
MG($N_c$,$|{\mathbb S}|$) with arbitrary strategy space ${\mathbb S}$ which
implies periodic dynamics does not always exist. Moreover, the periodicity
of the periodic dynamics in MG($N_c$,$|{\mathbb S}|$) (which is equal to the
periodicity of $C_0(i)$ as a function of $i$) is independent of the length
of memory $M$. In particular, we found that $C_0(i)$ is a period-$N_c^k$
function of $i$ in MG($N_c$,$N_c^k$) with $N_c > 2$ and $2 \le k \le 3$
whenever the prediction vector of each history is uniquely determined by one
of the following vectors alone: $\vec{\Lambda}(t-M)$,$\vec{\Lambda}(t-M-1)$,
$\ldots$, $\vec{\Lambda}(t-k+1)$ where  $\vec{\Lambda}(t) = (\Omega(t),
\ldots, \Omega(t+k-2))$. Besides, these periodic dynamics is observed to be
less and less pronounced for larger $N_c$ and $\kappa$ if the prediction
vector of each history is uniquely determined by $\vec{\Lambda}(t-\kappa)$
where the integer $\kappa \in [k-1, M]$.

To understand the dynamics in MG($N_c$,$|{\mathbb S}|$), let us start by
considering the dynamics of the game for every possible history. As
mentioned before, the virtual scores of all player's strategies must change
in the same manner for those histories with the same prediction vector and
thus the dynamics of the game for these histories cannot be considered
independently. We will consider the dynamics of the game for each possible
effective history instead in which those histories with the same prediction
vector are considered to be belonged to the same effective history. We
should note that there are totally $N_c^{k-1}$ distinct effective histories
in MG($N_c$,$N_c^k$) if the prediction vector only takes on $N_c^{k-1}$
different states for $|{\mathbb S}| = N_c^k$.

In fact, the effective history in MG($N_c$,$|{\mathbb S}|$) is analogous to
the history in the $N_c$-choice MG. Thus there should be a similar
period-$N_c$ dynamics for each effective history of
MG($N_c$,$|{\mathbb S}|$) when $\alpha$ is small. Our numerical findings
validates the existence of such periodic dynamics in
MG($N_c$,$|{\mathbb S}|$). However, do all effective histories have the same
probability of occurence in the overcrowding phase of
MG($N_c$,$|{\mathbb S}|$)? It is true only if the prediction vector of each
history is uniquely determined by one of the following vectors alone:
$\vec{\Lambda}(t-M)$, $\vec{\Lambda}(t-M-1)$, $\ldots$,
$\vec{\Lambda}(t-k+1)$. Under such circumstances, the current effective
history is determined by the corresponding past minority choice(s). In
addition, each alternative must have the same probability to be the minority
choice in the overcrowding phase under any history because there is no prior
bias for any alternative and all alternatives can be picked by some of the
players for such case. As a result, all effective histories must have the
same probability of occurence for small $\alpha$ in these
MG($N_c$,$|{\mathbb S}|$) games.

We revealed that there is a period-$N_c$ dynamics for all $N_c^{k-1}$
effective histories in the overcrowding phase of MG($N_c$,$N_c^k$) if the
prediction vector only takes on $N_c^{k-1}$ different states. Moreover, we
found that all effective histories on average appear once per $N_c^{k-1}$
time steps as they all have the same probability of occurence if the
prediction vector of each history is uniquely determined by
$\vec{\Lambda}(t-\kappa)$ where the integer $\kappa \in [k-1, M]$. Besides,
the probability for any transition of effective history must be the same for
such case which lead to no ``destructive interference'' between the dynamics
of individual effective history. As a results, there is period-$N_c^k$
dynamics in the overcrowding phase of MG($N_c$,$N_c^k$) under such
circumstances.

However, why the period-$N_c^k$ dynamics in MG($N_c$,$N_c^k$) becomes less
significant when $N_c$ increases? It is because the total number of possible
evolutionary path in the effective history space increases exponentially
with $N_c$ under the period-$N_c^k$ dynamics. Indeed, we have found that the
period-$N_c^k$ dynamics become more pronounced if we add bias to some of
the effective history's evolutionary path by manipulating the payoff
function. On the other hand, we should note that the evolution in the
effective history space depends only on $\Omega(t-\kappa), \ldots,
\Omega(t-1)$ if the prediction vector of each history is uniquely determined
by $\vec{\Lambda}(t-\kappa)$ where the integer $\kappa \in [k-1,M]$. Thus
the total number of possible evolutionary path in the effective history
space also increases exponentially with $\kappa$ and the periodic dynamics
become less pronounced when $\kappa$ increases.

Nevertheless, we found that $C_0(i)$ exhibits peculiar behavior in
MG($N_c$,$N_c^2$) with $N_c = 2$. In particular, $C_0(i)$ shows a
period-$N_c^3$ multi-peak structure whenever the prediction vector of each
history is uniquely determined by $\Omega(t-2\kappa)$ for any integer
$\kappa \in [1, \lfloor M/2 \rfloor]$. For MG($N_c$,$N_c^k$), the strategy
space is $k$-dimensional since the strategy space is spanned by $k$
independent strategies (see eqn.~(\ref{E:SS})-(\ref{E:Cond_uncorr})). Thus
the strategy space is 2-dimensional in MG($N_c$,$N_c^2$) which give rise to
the abnormal properties of $C_0(i)$. In addition, there are only 2 choices
for $N_c = 2$ which is remarkably different from those games with more than
two choices. It is because if a strategy does not choose room-0, then it
must choose room-1 and vice-versa for a 2-choice game. We believe the above
two factors lead to the marked difference of the dynamics of the system in
MG($N_c$,$N_c^2$) with $N_c = 2$ compared to other MG($N_c$,$|{\mathbb S}|$)
games with $N_c > 2$.  

In summary, we found that the periodic dynamics only exists in the
overcrowding phase for particular reduced strategy spaces in the
MG($N_c$,$|{\mathbb S}|$) game. In contrast, the periodic dynamics always
appears in the overcrowding phase for the original MG where all player's
strategies are drawn from the full strategy space.   

\section{Conclusion}

In conclusion, we have successfully build up the MG($N_c$,$|{\mathbb S}|$)
model whenever $N_c$ is a prime power. It is a generalized MG model where
each players has $N_c \ge 2$ choices and all strategies are drawn from the
reduced strategy space that is consisted of uncorrelated and anti-correlated
strategies only. Although our MG($N_c$,$|{\mathbb S}|$) model is not exactly
same as the MG, its global cooperative behavior is similar to that of the MG
no matter what $S$, ${\mathbb S}$ and $M$ are. 

However, the dynamics of MG($N_c$,$|{\mathbb S}|$) model is remarkably 
different from that of the MG. For the MG($N_c$,$|{\mathbb S}|$), the
periodic dynamics only exists in the overcrowding phase for particular
reduced strategy spaces; on the contrary, the periodic dynamics always
appears in the overcrowding phase for the MG.

Indeed, we have only given explanation of the periodic dynamics in the
MG($N_c$,$N_c^k$) model for some of the reduced strategy spaces in which a
periodic dynamics exists. Thus it is instructive for us to find reasonable
explanation for the periodic dynamics in other cases.

\begin{acknowledgments}
We would like to thank K. H. Ho for his useful discussions and comments.
One of the author (HFC) is supported in part by the Outstanding Young
Researcher Award of the University of Hong Kong.
\end{acknowledgments}

\pagebreak

\end{document}